# A High-Resolution Future Wave Climate Projection for the Coastal Northwestern Atlantic

Adrean WEBB[1], Tomoya SHIMURA[2] and Nobuhito MORI[3]

[1]Project Assistant professor, DPRI, Kyoto University
(Gokasho, Uji, Kyoto 611-0011, Japan)
E-mail:adrean.webb@gmail.com
[2]Member of JSCE, Assistant professor, DPRI, Kyoto University
(Gokasho, Uji, Kyoto 611-0011, Japan)
E-mail: mori.nobuhito.8a@kyoto-u.ac.jp
[3]Member of JSCE, Associate professor, DPRI, Kyoto University
(Gokasho, Uji, Kyoto 611-0011, Japan)
E-mail: shimura.tomoya.2v@kyoto-u.ac.jp

A high-resolution wave climate projection for the northwestern Atlantic Ocean has been conducted to help assess possible regional impacts due to global climate change. The spectral wave model NOAA WAVEWATCH III is utilized with three coupled (two-way) grids to resolve the northwestern Atlantic and coastal southern and eastern USA at approximately 21 km and 7 km respectively, and covers the periods 1979–2003 (historic) and 2075–2099 (future). Hourly wind field forcings are provided by a high-resolution AGCM (MRI-AGCM 3.2S; 21 km) and allow for better modeling of large storm events (important for extreme event statistics). Climatological (25-year) comparisons between future and historical periods indicate significant wave heights will decrease in the northwestern Atlantic Ocean (-5.7 %) and Gulf of Mexico (-4.7 %) but increase in the Caribbean Sea (2.4 %). Comparisons also indicate that large changes in mean wave direction will occur in the Gulf of Mexico (5.0°), with the largest occurring west of the Florida peninsula (over 15°).

## 1. INTRODUCTION

Many coastal systems and low-lying communities in the northwestern Atlantic Ocean (both tropical and northern areas) will experience adverse impacts such as submergence, flooding, and coastal erosion due to future projected relative sea level rises. Higher waves and surges in these regions will also increase the probability that coastal sand barriers and dunes will be overwashed or breached and more energetic and/or frequent storms will exacerbate all of these effects. As such, regional climate projections are becoming crucially important for identifying the affected regions and helping communities plan mitigation and adaptation strategies for the changing risks.

There have been several studies examining projected change for the entire Northern Atlantic ocean basin. In Wang et al.[1], GCM-derived sea level pressure is used to make projected changes in significant wave height and the authors find decreases in midlatitudes and increases in the southwest (for the North Atlantic). There have also been several regional hindcast studies. In Appendini et al.[2], a regional climatology for the Gulf of Mexico and Caribbean Sea was generated using a 30-year wave hindcast with a 12 km resolution. Analysis of the wave tendencies in the climatology for the Gulf of Mexico suggests there will be an increase in both mean and extreme wave conditions.

However, there is no available regional wave climate projection from the US Eastern Coast to the Caribbean Sea. As such, a regional, high-resolution wave climate projection for the northwestern Atlantic has been conducted based on a global projection. The regional projection includes several novel features and will be used in this study to help assess impacts due to global climate change. It is hoped that comparisons with other differing wave climates (such as Japan in the Pacific) will help elucidate differences and improve future regional projections.

Here, details of the wave model, simulation, and validation are given in Section 2. Regional sea and nearshore results and analyses are presented in Section 3. And finally, a discussion follows in Section 4.

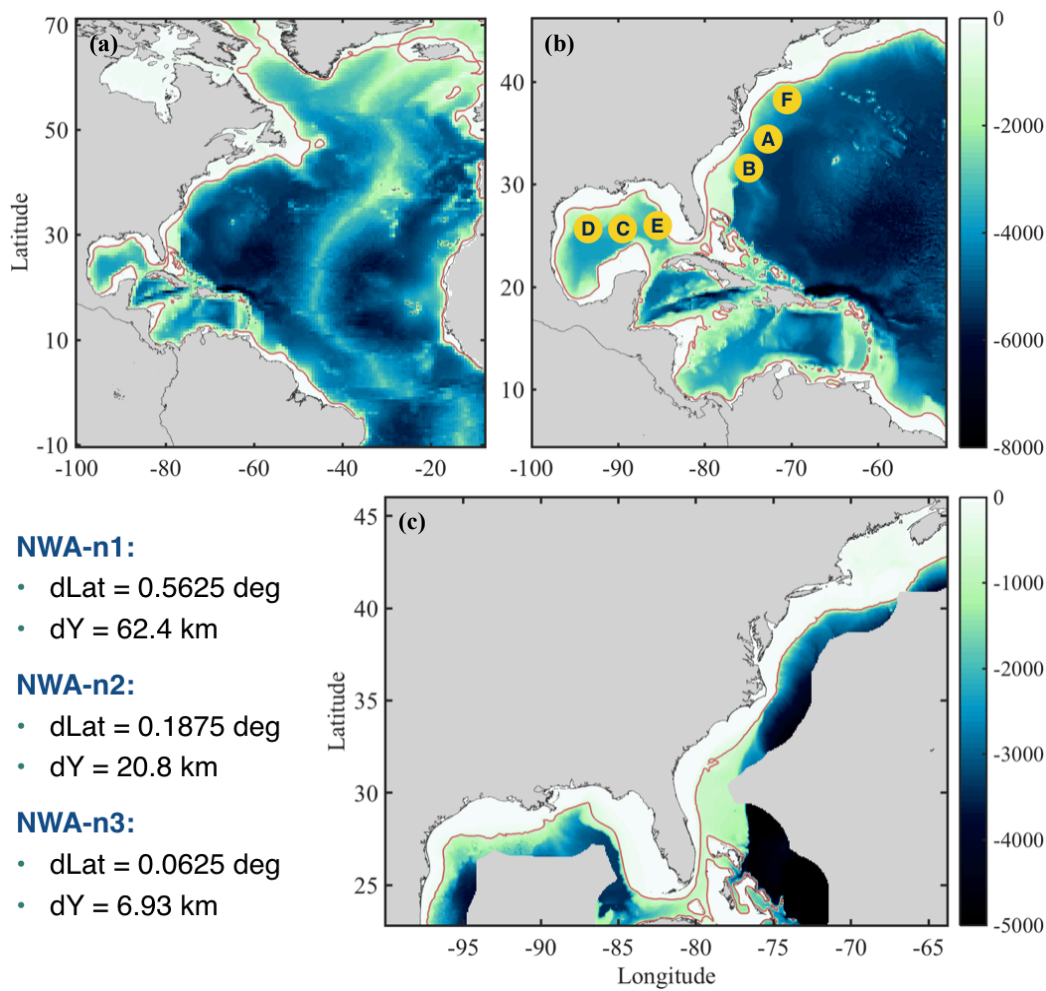

**Fig.1** NWA model bathymetries [m] for coupled grid setup (n1−3). NDBC buoy locations are shown in (b) and red contours indicate regions with 500 m depth.

## 2. MODEL SIMULATION

### (1) Model setup

The spectral wave model NOAA WAVEWATCH III (ver. 4.18) [3)] is utilized with three coupled (two-way) grids (n1−3) to resolve the northwestern Atlantic (NWA-n2) and US Southern and Eastern Coasts (NWA-n3) at approximately 21 km and 7 km, respectively (see **Fig.1**). The model has a spectral resolution of 29 frequency and 36 directional bins. It uses the Ardhuin et al. 2010 source term package (ST4, STAB0, and FLX0 with default $\beta_{max}$) and basic sea ice attenuation (IC0)—all other physics-based switches are default. The nested grids were generated using ETOPO1 (1/60°) bathymetry and GSHHS (ver. 2.3.4/5) shoreline data; details of the grids are listed in **Fig.1**.

### (2) Simulation details

An atmospheric general circulation model (AGCM) is a stand-alone atmospheric model that uses prescribed sea surface temperatures (SSTs) as a boundary condition. AGCMs have several benefits when used for climate studies: (a) higher horizontal spatial resolutions can be achieved and (b) present-day climatologies tend to be reproduced better since errors in observed SSTs tend to be smaller than those simulated by AOGCMs[4].

**Table 1** Details of datasets used for inter-model comparisons. ERA-I wave model is two-way coupled with its atmospheric counterpart.

| Wave model | Deg/lat | Output | Wind product | Deg/lat | Output | Source |
| --- | --- | --- | --- | --- | --- | --- |
| NWA-n1−3 | 62.4 / 20.8 / 6.9 | 1 hr | MRI-AGCM 3.2S | 20.8 | 1 hr | Kyoto Univ. |
| CFSR-n1−3 | 55.5 / 18.5 / 7.4 | 3 hr | CFSR | 34.6 | 1 hr | NCEP |
| JRA-55 | 62.4 | 1 hr | JRA-55 | 62.4 | 6 hr | Kyoto Univ. |
| ERA-I | 83.2 | 6 hr | ERA-Intrim | 77.9 | ** | ECMWF |

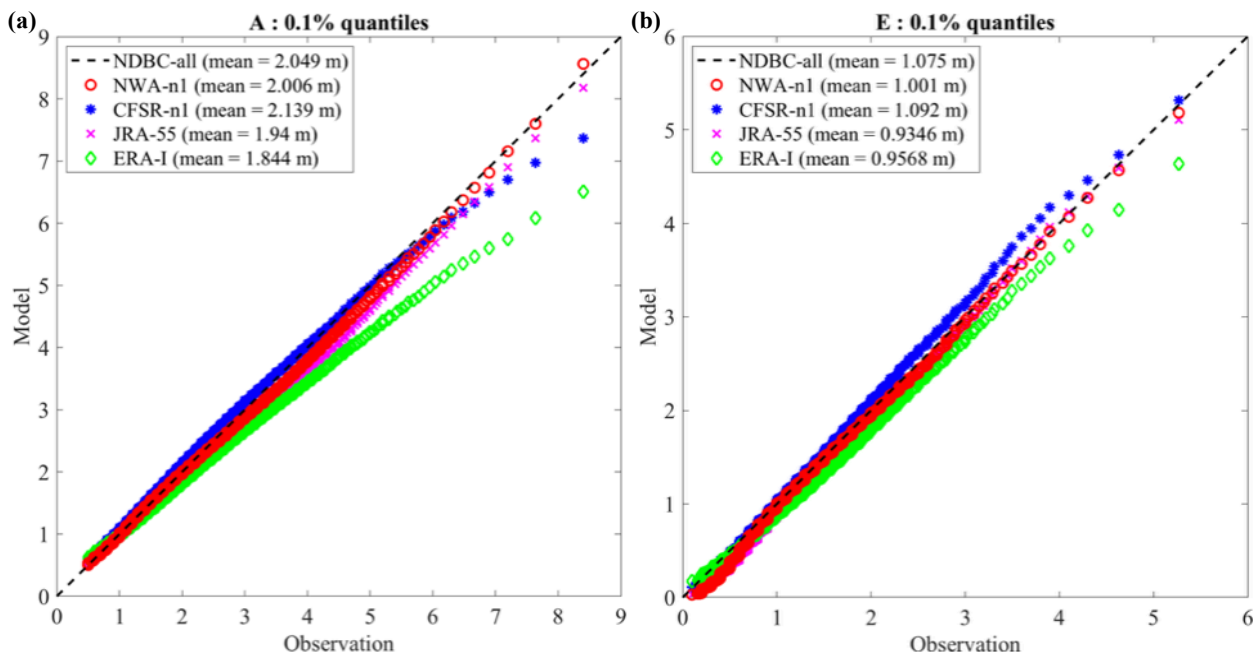

**Fig.2** Quantile-quantile plots (0.1%) of NDBC buoy records with historic period and three hindcast datasets at two locations.

Here, the main model input consists of hourly wind field forcings provided by the high-resolution MRI-AGCM 3.2S[5], with an approximate 21 km resolution. In addition, the Yoshimura cumulus convective scheme was used with an RCP8.5 climate forcing scenario. The simulations cover periods 1979–2003 (historic) and 2075–2099 (future) and provides hourly outputs.

**(3) Model validation and inter-model comparison**

NDBC buoy data were used to validate the wave climate model. In order to gauge the performance of the model, the buoy data was also compared with three hindcast wave datasets generated using different grid resolutions, atmospheric products, and/or wave models. For the analysis presented here, six locations are chosen; the locations of each buoy are indicated by the letters (a–f) in **Fig.1b**. The observational data used ranges from 31 to 38 years and consists both of 3 and 1 hour outputs. The three hindcast models used here are denoted as JRA-55, CFSR, and ERA-I and details of them are tabulated in **Table 1**. The CFSR model also consists of three nests, which are distinguished by the suffixes 'n1', 'n2', and 'n3'.

Quantile-quantile (qq) plots of significant wave height with 0.1% quantiles are shown in **Fig.2** for two selected buoys; for each analysis, all of the buoy data is compared with model output for 1979–2003. The results shown are similar at other locations and several patterns are evident. The NWA (wave) historic simulation tends to have the best agreement with the NDBC buoys (surprisingly), while the CFSR-n1 (wave) hindcast tends to overestimate and the ERA-I (wave) hindcast tends to underestimate. The JRA-55 atmospheric model uses the same dynamical core as the MRI-AGCM model (but different spatial resolution and output temporal frequency) and the JRA-55 (wave) hindcast has the closest agreement with the NWA (wave) historic simulation. Analyses were conducted for the higher resolution nests and similar results were found (not shown).

## 3. RESULTS

**(1) Regional sea analysis**

Climatological changes in wave height, period, and direction can have a big impact on coastal morphology. Here, 25-year climatologies have been calculated and analyzed for these variables for the historical and future periods. In **Fig.3a–c**, the future wave climate projections from NWA-n2 are shown for the spectral significant wave height ($H_{m0}$), first moment wave period ($T_{0,1}$), and mean wave direction ($\theta_w$); in **Fig.3d–f**, differences are shown between fu-ture and historic periods with relative differences using historic as reference. In addition, regional sea means and differences are tabulated in **Table 2** for the NWA-n2, Northwest Atlantic, Gulf of Mexico, and Caribbean Sea regions.

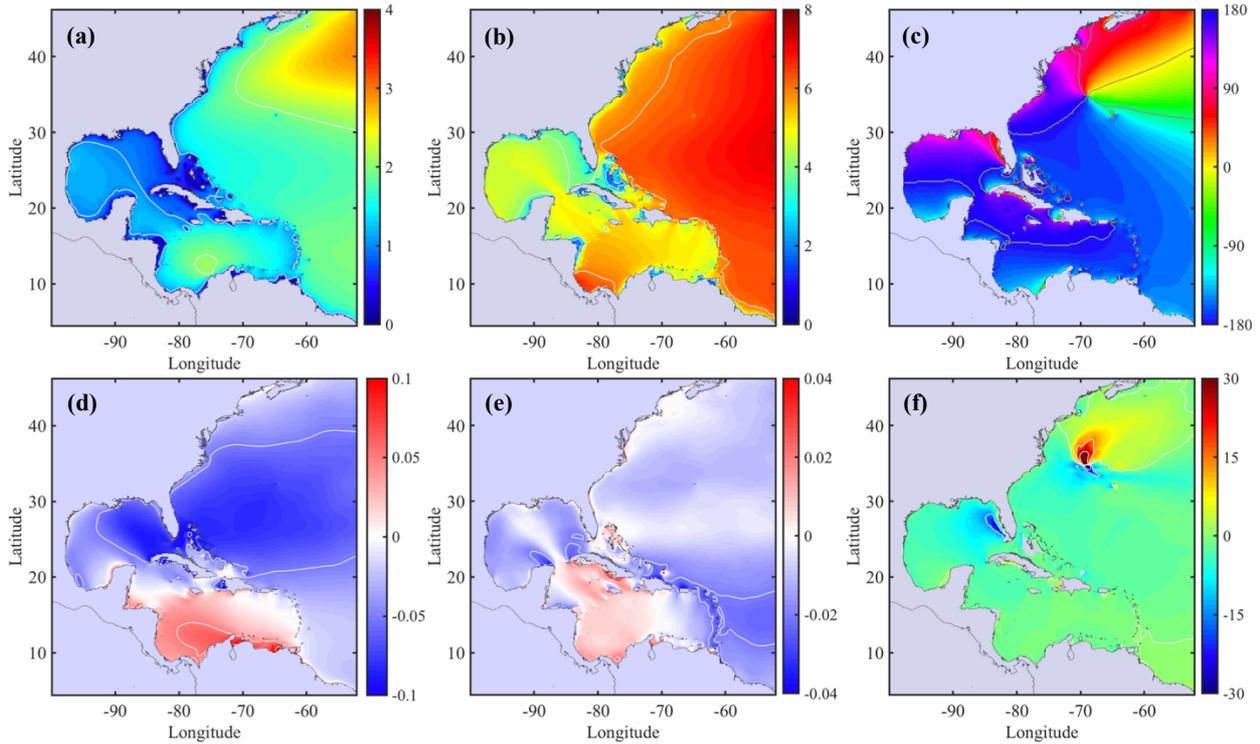

**Fig.3** 25-year climatologies of future wave climate projections (top row) and differences with historic period (bottom row). Analyses are shown for $H_{m0}$ [m; %] (left), $T_{01}$ [s; %] (center), and $\theta_w$ [deg; deg] (right); relative differences are shown for $H_{m0}$ and $T_{01}$ using historic as reference.

Examining $H_{m0}$, we see that there are two projected trends; there is an overall decrease in the Gulf of Mexico and northwestern Atlantic, and an overall increase in the Caribbean Sea, with relative differences -/+5% respectively in many parts of the regions. The largest mean decrease of -5.7% occurs in the Northwest Atlantic, whereas the largest mean increase of 2.4% occurs in Caribbean Sea. These changes in $H_{m0}$ coincide with changes in 10 m wind velocity magnitude ($U_{10}$); there are decreases above and increases below the 20° N latitudinal approximately, with relative differences roughly half to the same order (results omitted).

Mean wave direction patterns are fairly complicated in the NWA-n2 domain and vary greatly by location and season. Regional sea means are fairly consistent however and the overall climatological projections of $\theta_w$ for the NWA-n2 domain and each subdomain are approximately westward. Mean annual standard deviations for the historical and future periods are typically lowest (<15°) in the Caribbean Sea and highest (50° or more) west of the Florida peninsula and north of approximately 28° N latitude in the Atlantic Ocean (not shown). Differences between the future and historical $\theta_w$ climatologies range between 1.4° (Caribbean Sea) and 5.0° (Gulf of Mexico); similarly, large differences of over 15° are observed west of the Florida peninsula (**Fig.3f**).

**Table 2** Regional sea means of climatologies and differences. Relative differences (with historic used as reference) are tabulated for $U_{10}$, $H_{m0}$, and $T_{0,1}$; absolute (value) differences are tabulated for $\theta_w$.

| Variable | NWA-n2 | Northwest Atlantic | Gulf of Mexico | Caribbean Sea |
|---|---|---|---|---|
| $U_{10}$ (m/s) | 6.56 | 6.75 | 5.90 | 7.10 |
| $H_{m0}$ (m) | 1.46 | 1.73 | 0.97 | 1.38 |
| $T_{0,1}$ (s) | 5.37 | 6.04 | 4.30 | 4.99 |
| $\theta_w$ (deg) | -177 | -172 | 168 | -170 |
| $U_{10}$ (%) | -2.0 | -3.9 | -2.5 | 2.6 |
| $H_{m0}$ (%) | -3.7 | -5.7 | -4.7 | 2.4 |
| $T_{0,1}$ (%) | -0.7 | -0.7 | -0.9 | -0.1 |
| $\theta_w$ (deg) | 3.4 | 3.4 | 5.0 | 1.4 |

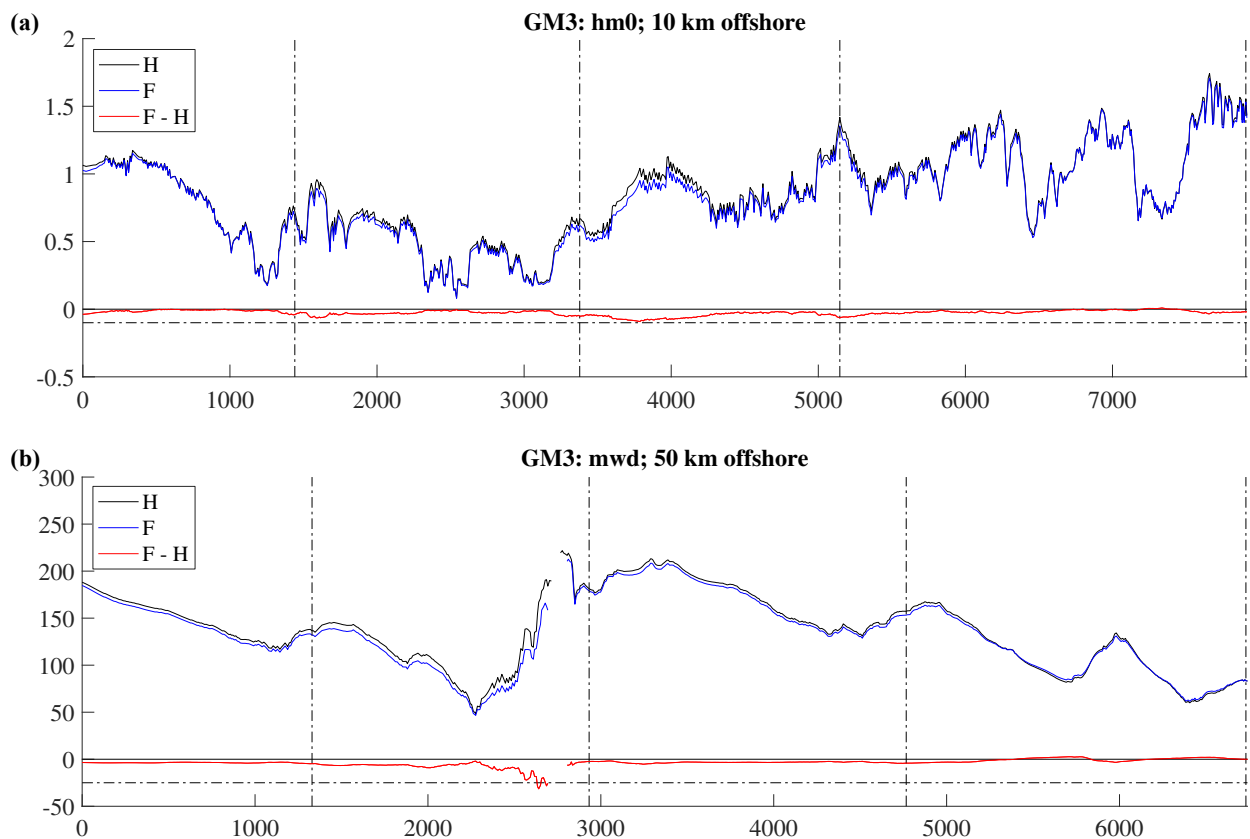

**Fig.4** Along-shore track analysis of (a) 10 km $H_{m0}$ [m] and (b) 50 km $\theta_w$ [deg].

The mean future climatologies of $T_{0,1}$ range approximately between 4 s and 6 s, and projected changes are much smaller, with relative differences less than -/+1% overall. The wave period can be used with the 10 m wind velocity magnitude to define a wave age as, $(g/2\pi) T_{0,1} / U_{10}$. The wave age climatologies are also smallest in the Caribbean Sea (which also corresponds to the smallest annual variations in $\theta_w$) and indicate fetch-limited conditions dominate as outside swell from the Atlantic is attenuated as it passes into the semi-enclosed basin (results omitted).

**(2) Coastal analysis**

To analyze nearshore changes, along-shore tracks were generated for the highest resolution grid (NWA-n3) at varying distances from the shore. In **Fig.4a–b**, historic and future climatologies and their differences are shown for $H_{m0}$ and $\theta_w$ at 10 km and 50 km respectively. The vertical lines represent three locations along the tracks: New Orleans (LA), Key West (FL), and Raleigh (NC).

The largest significant wave height peaks for both periods occur east of the Florida Keys (second vertical line), ranging from 0.5 m to 1.75 m approximately (**Fig.4a**). However, the largest differences roughly fall between New Orleans and Raleigh (first and third vertical lines), with decreases upwards of 10 cm. These differences tend to increase as the tracks are moved further offshore.

As previously mentioned, there are large changes west of the Florida peninsula (**Fig.4b**). The differences exceed 25° and tend to rotate northwards in a counter-clockwise fashion. These changes are notable and can have a significant impact on the coastal morphology, such as incident shoreline energy, and are worthy of further study.

## 4. DISCUSSION

Overall, the historic wave simulation performs well in comparison with other hindcast datasets and adds confidence to the projected changes in wave climatologies. For example, the increased temporal frequency of the atmospheric forcing allows for better modeling of large storm events, which is important for extreme event statistics in the Northwest Atlantic (such as in the higher quantiles in **Fig.2**). Common wind input frequencies (such as every 6 hr) are often too large for fast moving storms and are not spatially interpolated (in time) within the wave model correctly. To illustrate, a comparison of annual extreme wave heights for 2088 is shown in **Fig.5** using 1 hr and 6 hr outputs.

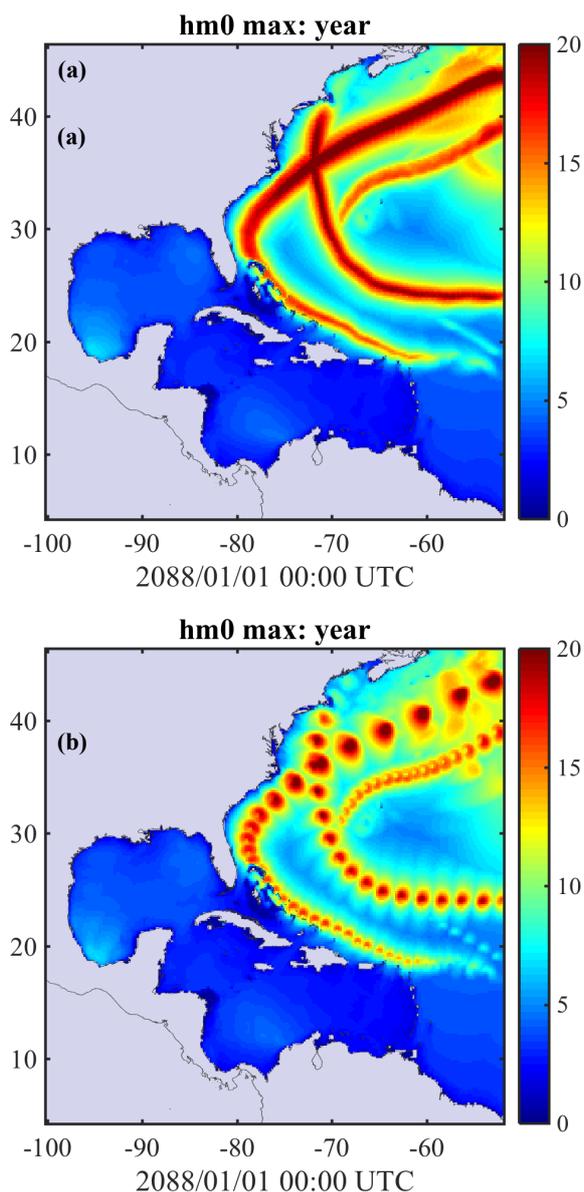

**Fig.5** Extreme NWA-n2 $H_{m0}$ for 2088 using different output frequencies [m]: (a) 1 hr and (b) 6 hr.

As previously mentioned, the coastal analyses presented tend to change depending on the distance of the along-shore track from shore; here, an increased spatial resolution allows for a more detailed projection of nearshore changes. The nearest possible tracks for the nested grids n1, n2, and n3 are approximately 3.5 km, 11.5 km, and 30 km from the shore respectively; depth effects can be substantial at these distances and projections with higher nearshore resolutions are a clear benefit.

Quantifying improvements further offshore (due to using a higher spatial resolution), such as in the regional sea analysis, is not as straight-forward since the nested domains are two-way coupled here and depth effects are less important. Differences between NWA-n1 and NWA-n2 regional sea means of the projected wave climatologies are small; they are within 0.2 m for $H_{m0}$, 0.04 s for $T_{0,1}$, and 1 deg for $\theta_w$. As a result, there are future plans to compare previous global wave climate studies with the regional study here in order to quantify other differences and improvements (if any).

**ACKNOWLEDGMENT:** This work was conducted under the framework of the Integrated Research Program for Advancing Climate Models (TOUGOU Program), supported by the Ministry of Education, Culture, Sports, Science, and Technology-Japan.